\documentclass[]{elsarticle}

\usepackage{lineno,hyperref}

\usepackage{natbib}
\usepackage{amssymb}
\usepackage{subfig}
\usepackage{graphicx}

\modulolinenumbers[5]

\journal{Icarus}

\biboptions{authoryear}

\begin{document}

\begin{frontmatter}


\title{Strength, stability and three dimensional structure of mean motion resonances in the Solar System}


\author{Tabar\'e Gallardo}
\address{Instituto de F\'{i}sica, Facultad
	de Ciencias, UdelaR, Igu\'{a} 4225, 11400 Montevideo, Uruguay}


\cortext[mycorrespondingauthor]{Corresponding author}
\ead{gallardo@fisica.edu.uy}


\begin{abstract}
In the framework of the circular restricted three body problem we show that the numerically computed strength $SR(e,i,\omega)$  is a good indicator of the strength and width of the mean-motion resonances in the full space $(e,i,\omega)$. We present a survey of strengths in the space $(e,i)$ for typical interior and exterior resonances. The resonance strength is highly dependent on $(e,i,\omega)$  except for exterior resonances of the type 1:$k$ for which the dependence with $(i,\omega)$ is softer. Such resonances are thus  strong even for retrograde orbits. All other resonances 
are weaker at very-high eccentricities for  $\omega\sim 90^{\circ}$ or $270^{\circ}$ and 
 $60^{\circ}\lesssim i \lesssim 120^{\circ}$.
 We explore the resonance structure in  the space $(a,i)$ by means of dynamical maps and we find structures similar to those of space $(a,e)$.
\end{abstract}


\end{frontmatter}


\section{Introduction}
\label{intro}

Orbital resonances are an essential 
mechanism 
 in the dynamics of minor bodies, planetary rings, satellite systems and planetary systems and they represent a fundamental core of knowledge of celestial mechanics. 
 In this paper we will focus on the case of a small body in mean motion resonance (hereafter MMR) with a planet,
 with the aim of extending our 
 understanding of its dynamics
towards regions of the space of orbital elements that have not yet been fully explored.
We recall that a particle with mean motion $n$ is in the MMR $k_p$:$k$ with a planet with mean motion $n_p$ when  the approximate relation  $k_p n_p - k n \sim 0$ is satisfied, being  $k_p$ and $k$ positive integers. The resonance is not limited to an exact value of  semimajor axis $a$, on the contrary the resonance has some width in astronomical units (au)
centered on the \textit{nominal} position, $a_0$, deduced from $n=n_p k_p/k$.
The picture astronomers have outlined along the years about resonant behavior is based, with few exceptions,  on  theories developed for low inclination orbits. These theories showed that the resonance domain in semimajor axis grows with the orbital eccentricity $e$: 
it goes from zero for $e=0$  to wide regions  for high $e$. In the case of the resonances with the giant planets of the Solar System, the resonant islands at high $e$ are so wide that a large chaotic region is formed, due to the superposition of the different resonances. There is a very complete  literature about MMRs, we can mention for example some chapters of books \citep{1999ssd..book.....M,2002mcma.book.....M,2007ASSL..345.....F,2010LNP...790....1L} and some reviews  \citep{1976ARA&A..14..215P,Malhotra98orbitalresonances,2002aste.conf..379N,2018P&SS..157...96G}.

\begin{figure}
	\centering
	\includegraphics[width=1.0\linewidth]{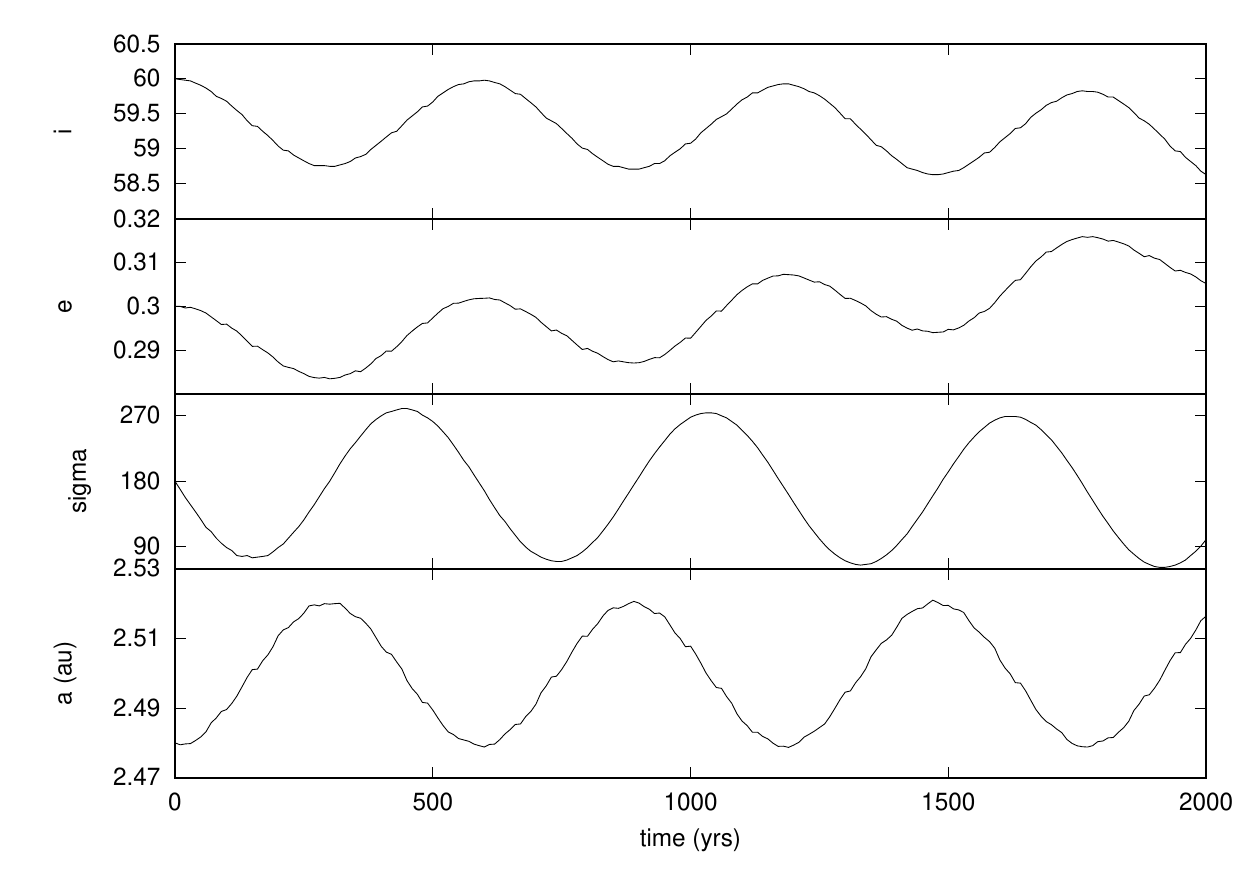}
	\caption{Time evolution of a test particle inside the resonance 3:1 with Jupiter.  The full set of orbital elements $(a,e,i)$ oscillates (or librates) due to the resonance. The critical angle is $\sigma = 3\lambda_J - \lambda - 2\varpi$. A drift much slower than the resonant oscillations is also clearly visible, as a result of the secular dynamics inside the resonance. These long-term variations can have a much larger amplitude for $e$ and $i$ than the small resonant oscillations, possibly leading the particle towards chaotic regions.}
	\label{fig:4plot}
\end{figure}

From basic theories, we know that the orbital dynamics of a small body in resonance with a planet
is defined by the disturbing function $R(a,\sigma)$, where $\sigma$ is the \textit{critical angle} that we will define later. The equations of motion can be derived from its Hamiltonian  $\mathcal{K}$, that can be found in the Appendix.
 The disturbing function $R$ actually depends also on the other orbital parameters of the small body, but their typical evolution timescale is generally much larger than $a$ and $\sigma$. All along this paper, we will focus only on the resonant (or semi-secular) timescale, over which $(e,i,\omega,\Omega)$ can be considered fixed.
The resonant motion imposes oscillations  (called \textit{librations}) of  $\sigma$ around an equilibrium value $\sigma_0$, correlated to oscillations of the semimajor axis $a$, though its value remains between limits defined by the borders (or \textit{separatrices}) of the resonance \citep{2002aste.conf..379N}. 
The interval between these limits is called \textit{width} of the resonance.
Simplified analytical theories based on a unique resonant perturbing term of the form $R = A\cos(\sigma)$ usually call \textit{strength}  the 
coefficient $A$. The simplified Hamiltonian adopts a pendulum-like form and then
the strength $A$ is thus equal to the depth of the resonance island, whereas its width is proportional to $\sqrt{A}$.
The overall geometry of the resonance is given by the level curves of $\mathcal{K}$ in the plane $(a,\sigma)$.
Of course, the remaining orbital elements $(e,i,\omega,\Omega)$ are actually not exactly fixed.
For example, we show
in figure \ref{fig:4plot} the time evolution of $a, e, i$ and $\sigma$ of a test particle evolving inside the 3:1 resonance with Jupiter. The pendulum-like oscillations of $a$ and  $\sigma$ are obvious.
Their repercussions on $e$ and $i$ are insignificant compared to their long-term drift (not shown and not studied in this paper):
we note in particular that the oscillations of $e$ and $i$ are exactly in phase with $a$, reflecting the fact that they are only a by-product of the coordinates used and not independent features of the dynamics.

The theories developed for low inclination orbits showed that in the low-eccentricity regime the strength of the resonance $k_p$:$k$ is  proportional to $e^{q}$ being $e$ the eccentricity of the particle and $q=|k_p-k|$. So, $q$ was conveniently called the \textit{order} of the resonance. This justifies that only low order resonances have deserved the attention of astronomers.
A complication to this simple rule was discovered by \cite{2012MNRAS.424...52M} and \cite{2013CeMDA.117..405M}. They demonstrated that for the extreme case of coplanar retrograde orbits (that means $i=180^{\circ}$) the strength of these  resonances is not proportional to the eccentricity elevated to the power $q$ but elevated to the power   $|k_p+k|$. Being these integers both positive the order for retrograde orbits results to be always larger. Then, the difference between the integers factorizing both $n$ is no longer representative of the order of the resonance for the full interval of orbital inclinations.
Recently an analytical expansion for near polar orbits was obtained  \citep{2017MNRAS.471.2097N} and it was found again a very different behavior: the expansion order of the disturbing function is not given by the value of $q$ but by its parity:  odd (1)  or even (2). That expansion was recently extended to arbitrary inclinations by \citet{2018MNRAS.474..157N}. Their paper lists the terms  up to fourth order terms in $e$ and $\sin(i-i_r)$ where $i_r$ is an arbitrary reference inclination.

In the general case, the leading-order terms of the disturbing function (including the so-called "pure eccentricity terms" of the classic expansions) are never proportional to $e$ alone,
but to coefficients of the type $e^N \sin i^M$, being $N$ and $M$ integers \citep{1998A&A...329..339R,2000Icar..147..129E,2018MNRAS.474..157N}. This generates complicated expressions.
 Any analytical representation of the disturbing function is accurate only in a restricted domain of the orbital parameters, and the number of terms with non-negligible strength increases dramatically as we get further from the reference value around which the disturbing function is expanded.
  Then $R(\sigma)$ cannot be more represented by an unique term but the concept of strength can be generalized to the amplitude of the exact $R(\sigma)$ which in this case must be calculated numerically \citep{2006Icar..184...29G}. Nevertheless, the concept of strength
  can still apply
   to a specific coefficient corresponding to some relevant critical angle as is done for example in \citet{2018MNRAS.474..157N}.

In numerical simulations of comets, centaurs and fictitious particles some works showed  that captures in retrograde resonances are a common orbital state
triggering the interest of the study of high inclination and retrograde resonances \citep{2015MNRAS.446.1998N,2016MNRAS.461.3075F,2018P&SS..158....6F}. In this context this paper  generalizes the concept of strength 
to the full range of orbital elements and facilitates its calculation by a numerical procedure.
We organize this paper as follows: in section \ref{stren} we  introduce the fundamental properties of the resonant motion, the numerical technique for computing the resonance strength, $SR$, for arbitrary resonances and we check $SR$ with the existing theories and with purely numerical methods, mainly dynamical maps.
 In section \ref{surv} we  present a survey of the strengths in the space $(e,i,\omega)$ for some typical resonances 
 still comparing the results to
 dynamical maps and we show some particular cases. In section \ref{strucai} we present the structure of MMRs in the space $(a,i)$. 
We summarize the conclusions in section \ref{conc}.

\section{Resonance strength}
\label{stren}

\subsection{Notation}

Different conventions have been utilized in the literature to describe the very simple relationship between the mean motions of two resonant objects. 
In this paper, we will call resonance $k_p$:$k$ the resonance generated by the commensurability  given by  $ k_p n_p - k n\sim 0$.
For example, 3:1 is a resonance interior to the perturbing planet and 1:3 is an exterior resonance. 
Following for example \citet{2000Icar..147..129E},  the resonant disturbing function, $R(\sigma)$, 
can be written as a series expansion of cosines which arguments are of the type
\begin{equation}\label{sigma}
\sigma =  k_p \lambda_p - k \lambda  + \gamma
\end{equation}
where $\lambda_p$ and $\lambda$ are the quick varying mean longitudes of the planet and particle respectively and $\gamma$ is 
a slow evolving angle defined by a linear combination of the longitudes of the ascending nodes and longitudes of the perihelia of the particle and the planet involved.
 In the simplified case of a perturbing planet with zero inclination and circular orbit   $\gamma$ only depends on the asteroid's  longitude of perihelion $\varpi$ and longitude of the ascending node $\Omega$ \citep{2006Icar..184...29G,2013CeMDA.117..405M}. Different linear combinations of $\varpi$ and $\Omega$ generate different $\gamma$ and consequently different $\sigma$, but all of them include the angle $k_p \lambda_p - k \lambda$, characteristic of the resonance $k_p$:$k$.
 All possible $\sigma$ can be called critical angle but in general there is one particular $\sigma$ that correlates better with the 
 oscillations of $a$. That $\sigma$ in general is the argument of the most relevant term in the expansion of the resonant disturbing function. By relevant, we mean that the coefficient factorizing the term containing $\cos(\sigma)$ is the largest and in consequence, it dominates the resonant dynamics.
For example, for low-inclination and low-eccentricity orbits, in general the critical angle that describes better the resonant motion is the one defined by $\gamma = (k - k_p)\varpi$ (still considering a perturbing planet with zero $e$ and $i$).
For resonances involving  high-eccentricity or/and high-inclination orbits there are several terms in the disturbing function that we must take into account and we cannot describe with a unique critical angle the complexity of the resonant dynamics.

\subsection{Numerical computation of the resonance strength, $SR$}

For a quick estimation of the resonance strength we have proposed \citep{2006Icar..184...29G} a numerical method that gives the semi-amplitude (called $SR$) of the resonant disturbing function for a given resonance with a given planet assumed in a zero inclination and circular orbit. 
In that approximation, for the purpose of the numerical computation of  $R(\sigma)$, we assume fixed orbits for both the particle and the planet, taking for the particle the semi-major axis corresponding to the nominal position of the resonance. 
This is justified by the slow evolution timescale of $(e,i,\omega,\Omega)$ as compared to the oscillations of $a$ and $\sigma$. Similarly, the evolution of $a$
and  $\sigma$ is slow compared to the orbital periods.
In this case, as the planet is in  a zero inclination and circular orbit it results that 
$SR$ is independent of $\Omega$ and it
 follows that for a given resonance $SR$ only depends on $(e,i,\omega)$.
Details of the numerical calculation of $R(\sigma)$ can be found in the Appendix.
Minima of  $R(\sigma)$ correspond to the stable equilibrium values for $\sigma$. For example
the minimum of the $R(\sigma)$ corresponding to the resonance showed in figure \ref{fig:4plot} occurs for $\sigma=180^{\circ}$.
Sometimes it is possible that  $R(\sigma)$ tends to a very large value for a specific value  $\sigma_{\infty}$ because in that case a close encounter with the planet occurs. A resonant object can evolve under the perturbation of a divergent  $R(\sigma)$ as long as the object does not reach  $\sigma \sim \sigma_{\infty}$, otherwise the resonant motion will be destroyed. 
The idea of computing numerically $R(\sigma)$ can be traced back to \citet{1968AJ.....73...99S}.

The critical angle can have different definitions according to  direct or retrograde orbits \citep{2012MNRAS.424...52M,2018MNRAS.474..157N} but in the calculation of $SR$ we consider only the classic definition for direct orbits and
this does not affect the results for $SR$ because it is a numerical estimation of the semiamplitude of the full disturbing function and not an estimation of the coefficient corresponding to the term associated to a specific critical angle. Then, different definitions of the critical angle affect the location of the maxima and minima of the corresponding 
$R(\sigma)$
 but not their value nor the semiamplitude of $R(\sigma)$. In fact, if $\sigma$ and $\sigma'$ are two different critical angles for the same resonance, we have $R(\sigma) = R(\sigma' + C)$ being $C$ a constant. So, $SR$ is independent of the definition of the critical angle provided $k$ and $k_p$ are the same. In other words, $SR$ only depends on $\sigma$ through the term $k_p \lambda_p - k \lambda$ which is the same for all critical angles of the resonance $k_p$:$k$.
Let us note that our $R(\sigma)$ is not a particular term of the resonant disturbing function, it is a numerical representation of the full disturbing function.

Values of $SR$ obtained with this method are 
exact and do not rely on series expansions. They are thus valid in the entire range of orbital elements (and they fully agree with series expansions in their respective ranges of validity).
In particular, the method is also valid for retrograde orbits and high eccentricity orbits as we will show later.
It is possible to correlate the strength $SR$ with the width of the resonance in au as has been done by  \citet{2011MNRAS.414.1059S} for a particular case of meteoroids streams at low eccentricity regime. At the end of this section we will show that the width is approximately proportional to $\sqrt{SR}$ even for high eccentricity and high inclination orbits.
Also the strength $SR$ and the \textit{stickiness} can be correlated as was first studied systematically by \citet{2007Icar..192..238L} in the transneptunian region.
The \textit{average time lead/lag}, $<dtr>$, was introduced by \citet{2016ApJ...816L..31M}  and it is a measure of the effect of the  resonance on the drift in $a$ generated by, for example, the Yarkovsky effect. They also found a correlation between  $<dtr>$ and $SR$.

The method for calculating $SR$ rely on the constancy of both orbits: particle and planet. The planet can be considered in an arbitrary $(e,i)$-orbit for the calculation of $SR$ but it is essential that in an interval of time comparable with the libration period both orbits do not change appreciably otherwise the resonance strength
will not be well represented
by $SR$. It is important to stress that once a particle is inside a MMR it could be subject to secular evolutions 
\citep{2002mcma.book.....M}
that this method is unable to predict. 
The codes for the numerical computation of the resonant disturbing function and $SR$ can be found at www.fisica.edu.uy/$\sim$gallardo/atlas/.

\subsection{Testing $SR(e)$ for direct, retrograde and polar orbits}
\label{eva}

\begin{figure}
	\centering
	\includegraphics[width=0.9\linewidth]{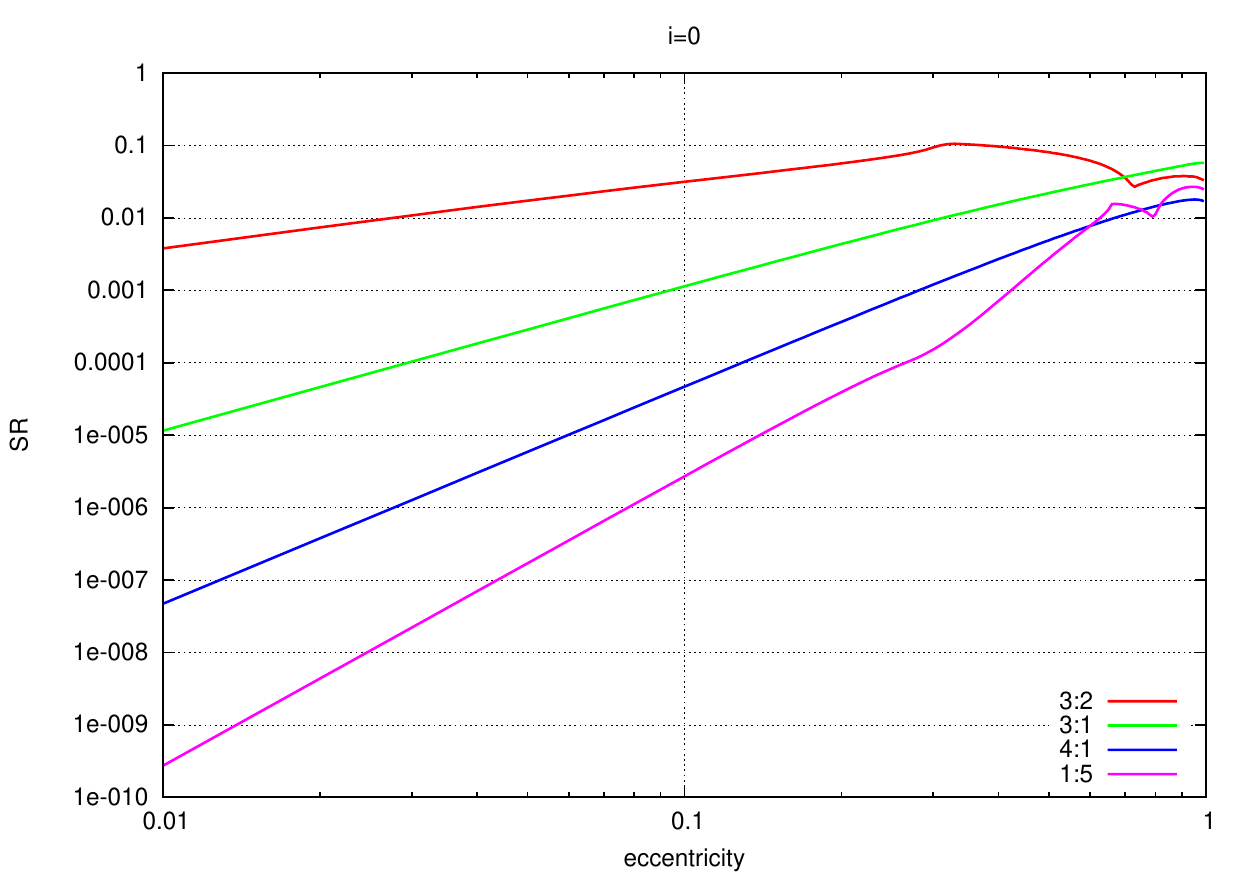}
	\caption{$SR(e,i=0^{\circ})$ in log-log scale for four resonances computed for nominal position in semi-major axis. The value of $\omega$ is irrelevant for zero inclination orbits. In the low-eccentricity regime the interior resonances 3:2, 3:1 and 4:1  go with $e$, $e^2$ and $e^3$ respectively and the exterior resonance 1:5 goes with $e^4$. At high eccentricities this simple trend is not more valid.}
	\label{fig:inc0}
\end{figure}

\begin{figure}
	\centering
	\includegraphics[width=0.9\linewidth]{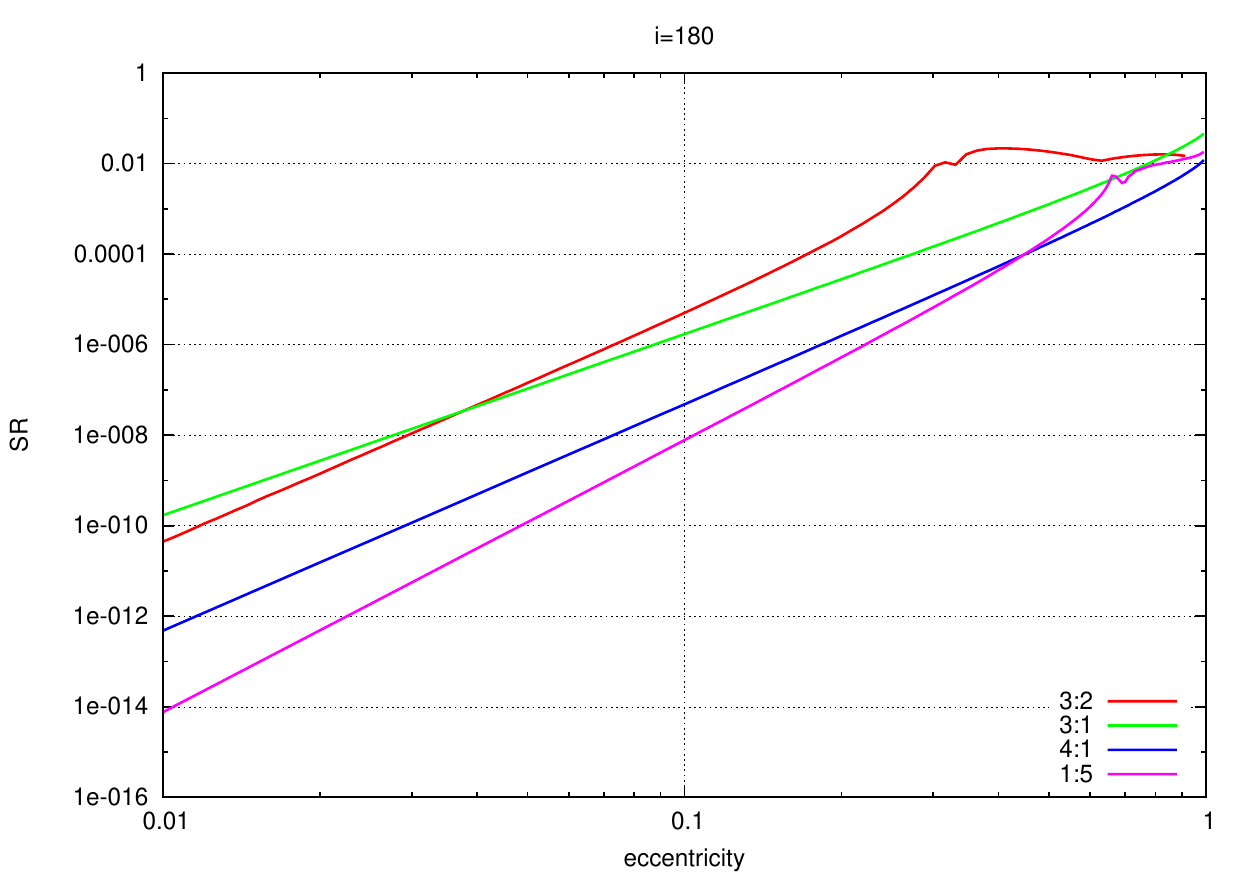}
	\caption{$SR(e,i=180^{\circ})$ for the same four resonances of figure \ref{fig:inc0}. In the low-eccentricity regime the resonance 3:1 goes with $e^4$, resonances 3:2 and 4:1 go with $e^5$ and 1:5 goes with $e^6$.}
	\label{fig:inc180}
\end{figure}

We show here the  strengths given  by the function   $SR(e)$ for given values of  $i$ and $\omega$ for various resonances and we check with predictions by theoretical models. Part of these tests can be found in \citet{2006Icar..184...29G} but we will extend them to wider regions in the space $(e,i,\omega)$.
Following that paper, we also present the results divided by $G m_J$, being $G$ the gravitational constant and $m_J$ the mass of Jupiter, in order to facilitate the reading of the plots.
We show in figure \ref{fig:inc0} the computed $SR(e,i=0^{\circ})$ for resonances of order 1, 2, 3 and 4 where it can be corroborated that $SR\propto e^{|k_p -k|}$ in the low eccentricity regime as theories predict. 
For example, for the resonance 3:2 $SR$ increases an order of magnitude when $e$ goes from 0.01 to 0.1 and
 for the resonance 1:5 $SR$ increases four orders of magnitude in the same interval variation of $e$.
In figure  \ref{fig:inc180} we show  $SR(e,i=180^{\circ})$ for the same resonances showing that $SR\propto e^{|k_p+k|}$ in the low eccentricity regime   confirming the predictions by \cite{2012MNRAS.424...52M} and \cite{2013CeMDA.117..405M}.
These behaviors of $SR(e)$ for coplanar orbits in the low eccentricity regime are explained 
because one term of the development largely dominates the others and we can safely take it alone as a first approximation
 in the analytical expansion of the resonant disturbing function. For high eccentricities the dependence of $SR$ with $e$ is more complex than a simple power of $e$ because there are several terms contributing to $R(\sigma)$.

\begin{figure}
	\centering
	\includegraphics[width=0.9\linewidth]{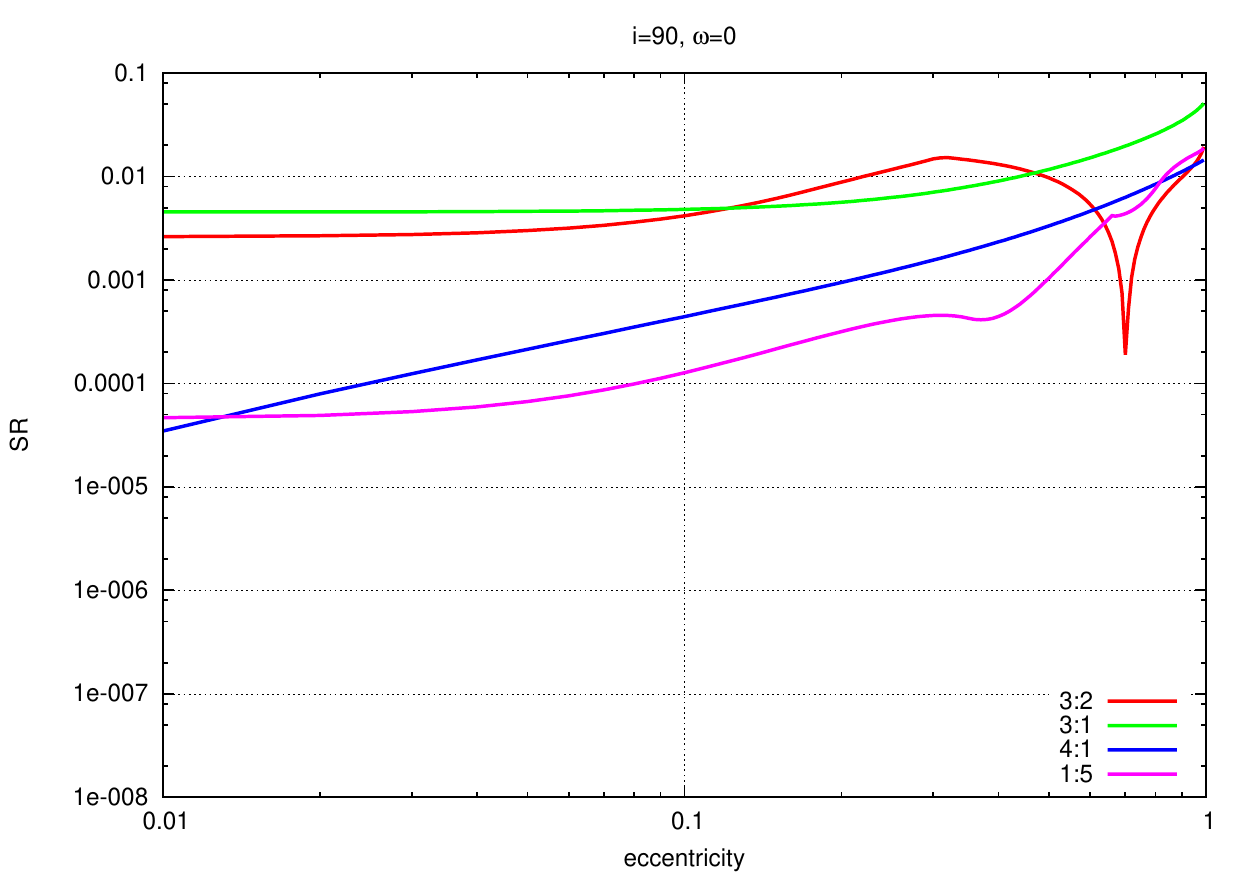}
	\caption{$SR(e,i=90^{\circ},\omega=0^{\circ})$ for the same four resonances of figure \ref{fig:inc0}. In the low-eccentricity regime the resonance 4:1 goes with $e$, 3:1 is independent of $e$ and the others have a more complex dependence with $e$.}
	\label{fig:inc90}
\end{figure}

\begin{figure}
	\centering
	\includegraphics[width=0.9\linewidth]{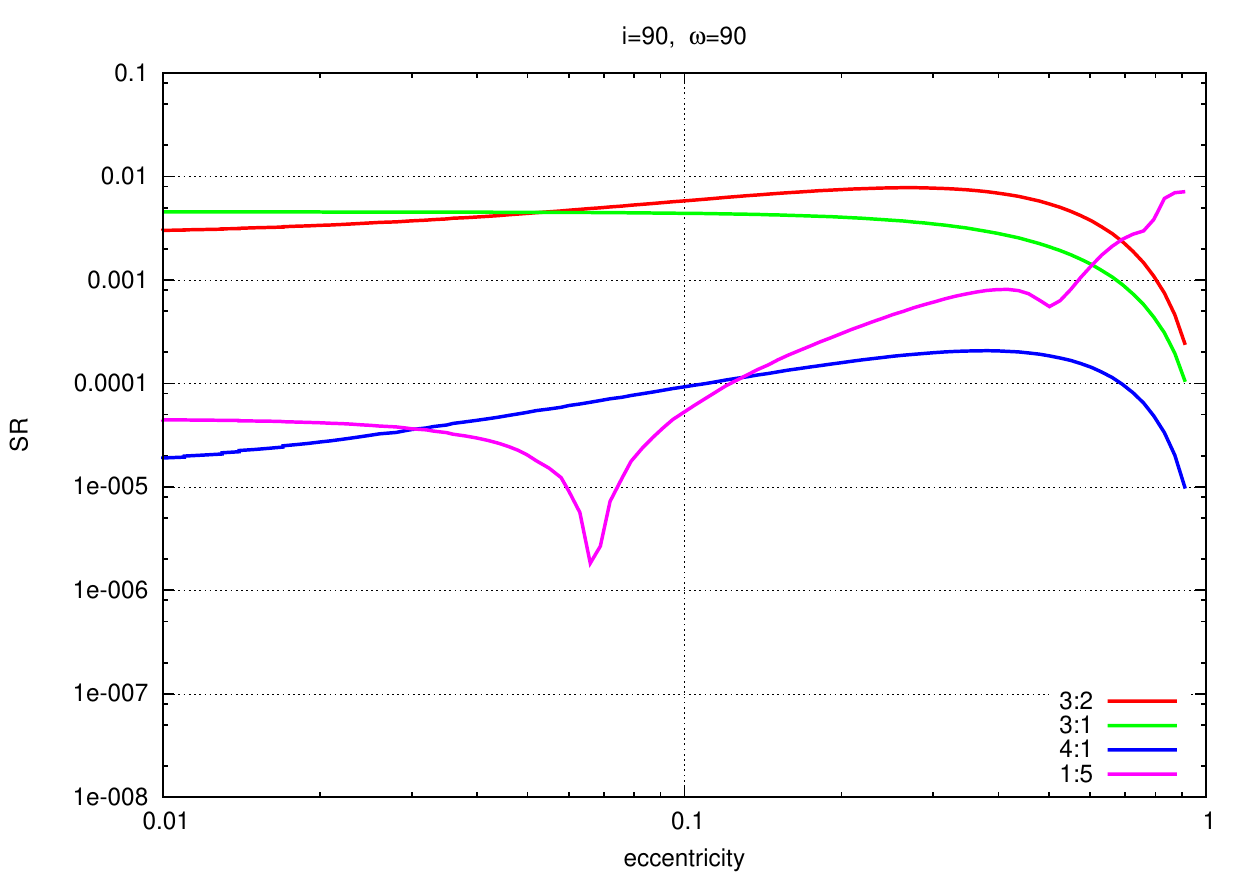}
	\caption{$SR(e,i=90^{\circ},\omega=90^{\circ})$ for the same four resonances of figure \ref{fig:inc90} showing a complex dependence with $e$.  Note that for interior resonances when $e\rightarrow 1$, $SR\rightarrow 0$.}
	\label{fig:inc90w90}
\end{figure}

In the case of non coplanar orbits there are various terms depending on $(e,i)$ in the disturbing function and the analytical 
approximation of $SR(e,i,\omega)$ based on series expansion is not trivial.
For example we show here the particular case of polar orbits ($i=90^{\circ}$).
 In figure \ref{fig:inc90} we show   $SR(e,i=90^{\circ},\omega=0^{\circ})$ for the same resonances of the previous figures and in figure \ref{fig:inc90w90} we show  $SR(e,i=90^{\circ},\omega=90^{\circ})$. From the analysis of these figures we can conclude: first, 
   the value of the argument of the perihelion becomes relevant
   and second,
    the 
   behavior of   $SR$ as a function of $e$ is not as simple as in the coplanar case.
 We remark that for resonant polar orbits the dynamics of resonances depend strongly on $\omega$, except when $e \sim 0$, where $\omega$ becomes irrelevant as can be confirmed in figures \ref{fig:inc90} and \ref{fig:inc90w90} for $e \sim 0$.
 For example, 
according to our results for polar orbits the strength of the resonance 3:1 seems to be independent of the eccentricity for $e<0.2$ but the following behavior strongly depends on $\omega$: the strength increases in the case $\omega=0^{\circ}$ (figure \ref{fig:inc90}) and a decreases in the case of $\omega=90^{\circ}$  (figure \ref{fig:inc90w90}). 
This is in contradiction with the classic idea: here, for higher eccentricities, the resonance is weaker! We will  come back to this issue later.

The dependence of $SR$ on $e$ for polar orbits that we have obtained 
seems to have some discrepancy with the affirmation given in  \citet{2017MNRAS.471.2097N} where they stated that all even resonances have strengths proportional to $e^2$ and all odd resonances have strengths proportional to  $e$. The reason is that we are calling strength to the total effects of all involved terms while in \citet{2017MNRAS.471.2097N} the strength is the coefficient associated to a specific critical angle. This discrepancy shows that, whereas the leading-order term of the expansions by \citet{2017MNRAS.471.2097N} is indeed proportional to $e$ or $e^2$, the remaining terms play a quite substantial role and cannot be neglected in general.

\subsection{Testing $SR$ by means of dynamical maps}
\label{map}

In order to corroborate our predictions about resonance strength obtained by means of $SR$  we constructed dynamical maps for some resonances following the technique described in \citet{2016Icar..274...83G}.
In this case we numerically integrate test particles with initial conditions taken
in a grid of some interval of  $a$ and $e$ (or $i$) for a fixed value of $i$ (or $e$) and with particular initial value for $\omega$.
We considered Jupiter as the only perturbing planet in a circular and zero inclination orbit. We integrated for some time interval long enough for covering a few libration periods using 
 a code which is a modification of Orbe \citep{2017EJPh...38c5002G} with a constant timestep of approximately the 1/40 part of the minimum orbital period of the intervening bodies.
We eliminate the short period oscillations of the computed $a,e,i$  calculating mean values in intervals of some tens of years, that means short intervals in comparison with the libration period. We then calculate the interval of variation of the mean elements $\Delta a$, $\Delta e$ and $\Delta i$.
In the middle of the resonance these $\Delta$ are minimal because the test particles are integrated with initial conditions exactly at the center where the equilibrium points are located. In the borders of the resonance the $\Delta$ are maximal because they correspond to particles  with initial conditions near the separatrices. Outside the resonance  the $\Delta$ are not null but drop abruptly. 
This produces a kind of bifurcation diagram, but directly obtained from purely numerical integrations without any assumption.
With this kind of map we can obtain a measure of the width of the resonance, that means the distance between the separatrices, and its dependence with $(e,i,\omega)$ and we can compare with our $SR(e,i,\omega)$. It is important to take into account that different time intervals for the numerical integrations can generate some variations in the absolute values of the plotted  $\Delta a$  but the structures persist.

\begin{figure}
	\centering
	\includegraphics[width=1.\linewidth]{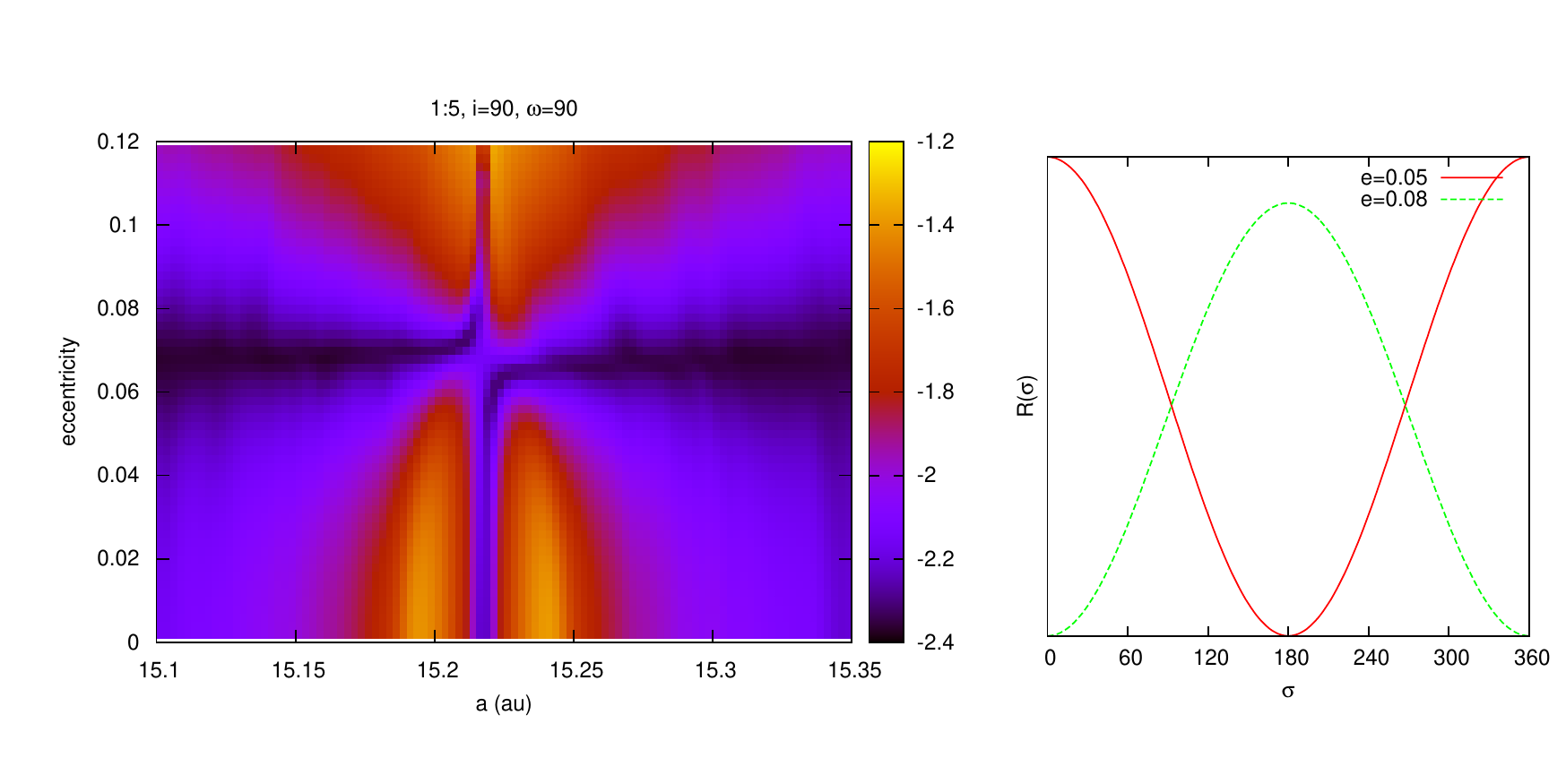}
	\caption{Left: dynamical map for the exterior resonance 1:5 with Jupiter for initial $i=90^{\circ},\omega=90^{\circ}$ constructed from a series of  numerical integrations of 4000 yrs. Color code indicates $\log(\Delta a)$ in au. The center of the resonance is  at $a=15.22$ au and the borders are approximately located in the yellow-orange regions.
	There is an evident break around $e \sim 0.07$ which also appears in figure \ref{fig:inc90w90}. Right: the corresponding $R(\sigma)$ for this resonance evaluated for $e = 0.05$ and
 $e = 0.08$ showing that the break in resonance strength at  $e \sim 0.07$ is due to a change in the stability of the equilibrium centers. }
	\label{map1to5}
\end{figure}

 For example, in the left panel of figure \ref{map1to5} we show a dynamical map $\Delta a(a,e)$ for initial values $i=90^{\circ},\omega=90^{\circ}$ for the region of the exterior resonance 1:5 where, according to figure \ref{fig:inc90w90}, $SR$
drops at  $e \sim 0.07$. The initial conditions were taken satisfying $\sigma =\lambda_J - 5\lambda + 4\varpi = 180^{\circ}$  which is the location of the equilibrium point for that critical angle for $e\sim 0$ according to the disturbing function calculated following the procedure indicated in the Appendix. 
The scale of colors corresponds to the logarithm of  $\Delta a$ representing with vivid colors larger  $\Delta a$ and with dark colors smaller  $\Delta a$.
The center of the resonance is clearly defined in the map at $a=15.22$ au where the variations  $\Delta a$ are minimal. 
The dynamical map 
 in figure \ref{map1to5} 
shows that for $e<0.05$ the width, roughly defined by the yellow and orange zone, is almost constant and that for  $e \sim 0.07$ the
width shrinks and 
this confirms that the resonance width follows the same variations as its strength $SR$
showed in figure  \ref{fig:inc90w90}. For  $e > 0.07$ the width of the resonance grows in agreement with the trend showed by  $SR$
in figure \ref{fig:inc90w90}. The disruption of the resonance at $e \sim 0.07$ is due to a change in the stability of the equilibrium points: for $e < 0.07$ $\sigma_0 = 180^{\circ}$ is stable and for $e > 0.07$ it becomes unstable as is showed in the right panel of figure  \ref{map1to5}.

\begin{figure}
	\centering
	\includegraphics[width=1.\linewidth]{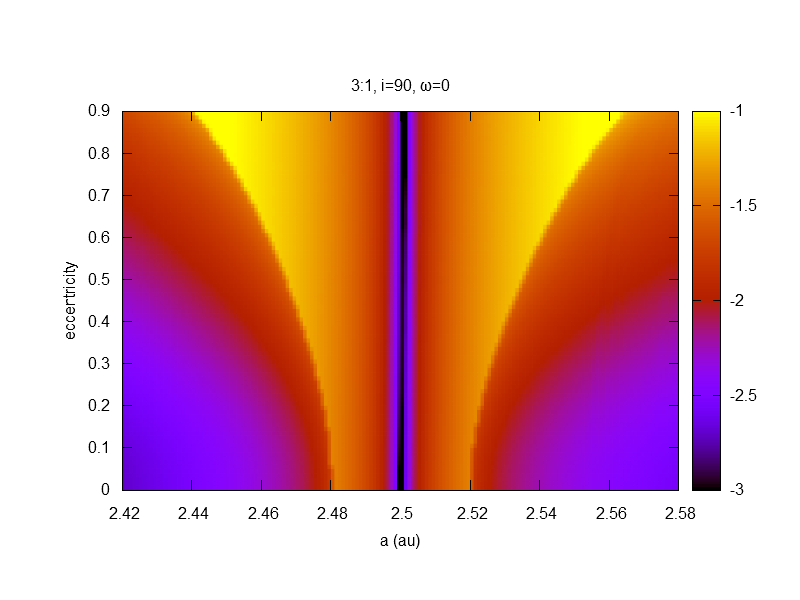}
	\caption{Dynamical map for the resonance 3:1 for initial $i=90^{\circ},\omega=0^{\circ}$. Code color indicates $\log(\Delta a)$ in au where $a$ is a mean value calculated in an interval of 20 yrs in order to eliminate short period terms. $\Delta a$ is the interval of variation detected in 1000 yrs. The width of the resonance is approximately constant up to $e\sim 0.2$ and then it grows for growing $e$. Compare with figure \ref{fig:inc90}.}
	\label{fig:3to1i90w0}
\end{figure}

\begin{figure}
	\centering
	\includegraphics[width=1.\linewidth]{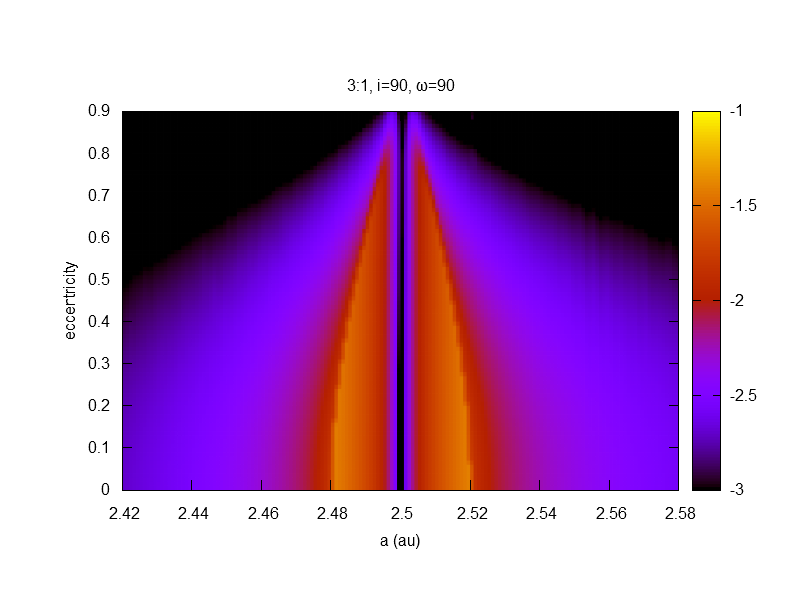}
	\caption{Same as figure \ref{fig:3to1i90w0} but for initial $\omega=90^{\circ}$.  The width of the resonance is approximately constant up to $e\sim 0.2$ and then, contrary to the classical view, the width of the resonance diminishes for growing $e$. Compare with figure \ref{fig:inc90w90}.}
	\label{fig:3to1i90w90}
\end{figure}

The uncommon behavior of polar resonances showed in
 figures \ref{fig:inc90} and \ref{fig:inc90w90} can also be checked with the dynamical maps 
 $\Delta a(a,e)$ for the resonance 3:1  calculated  for $i=90^{\circ}$ and two values of $\omega$, $0^{\circ}$ and $90^{\circ}$, which are showed
 in figures \ref{fig:3to1i90w0} and  \ref{fig:3to1i90w90}.
The dark vertical line in these figures is due to almost null oscillations at the center of the resonance
and the sharp edges at both sides are large amplitude librations near the separatrices marking the limits of the resonance. 
These figures confirmed what we have predicted in figures \ref{fig:inc90} and \ref{fig:inc90w90} for this resonance: its strength is almost constant for  $e<0.2$ and then the width increases in the case $\omega=0^{\circ}$ and decreases for $\omega=90^{\circ}$, vanishing for very high eccentricities.
It is evident that the argument of the perihelion is relevant for defining the resonance strength and width. It is  worth noting that changes in the topology of the resonance due to $\omega$ generate in turn discontinuities in the long-term secular evolution of resonant minor bodies as pointed out by \citet{2016CeMDA.126..369S}. 

\begin{figure}
	\centering
	\includegraphics[width=1.\linewidth]{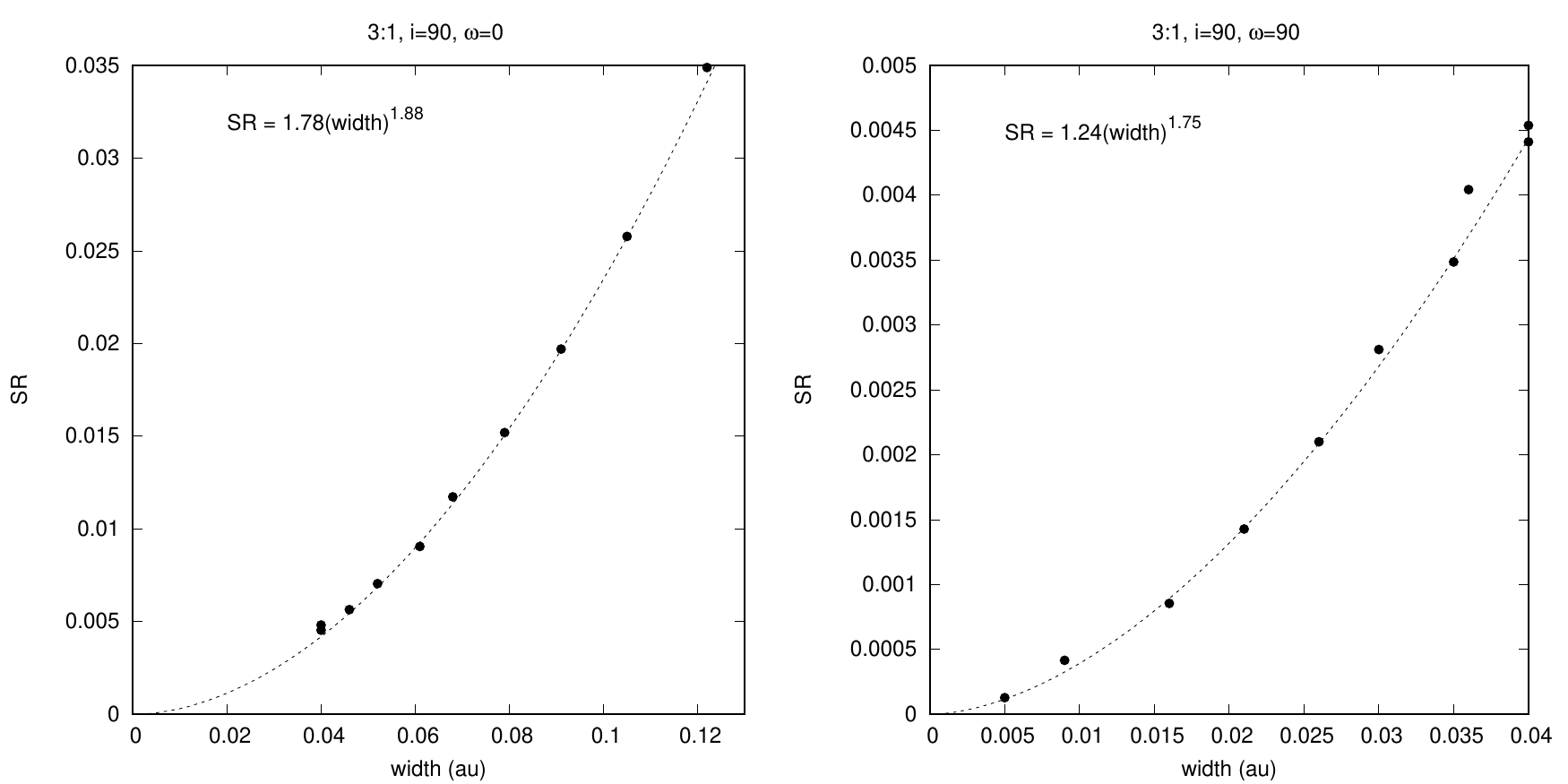}
	\caption{Left panel: correlation between $SR$ of the resonance 3:1 calculated from figure \ref{fig:inc90} and widths measured from figure \ref{fig:3to1i90w0}. Right panel: correlation between $SR$ of the resonance 3:1 calculated from figure \ref{fig:inc90w90} and widths measured from figure \ref{fig:3to1i90w90}. }
	\label{srwidth}
\end{figure}

We investigated the relation between $SR$ and the width of the resonance 3:1 obtained by direct inspection of figures \ref{fig:3to1i90w0} and  \ref{fig:3to1i90w90} and comparing with the corresponding curves in figures \ref{fig:inc90} and  \ref{fig:inc90w90}. Figure \ref{srwidth} shows the data used and fitting curves obtained. For these two cases we found that $\Delta a \propto SR^{0.55}$ which is very close the theoretical result $\Delta a \propto \sqrt{SR}$ obtained from simple pendulum-like resonant disturbing functions.  We can conclude that $SR(e,i,\omega)$ is a confident indicator of the full strength and width of the resonances, even for retrograde orbits and/or very high eccentricity orbits, and it can be used to explore the resonances in all range of orbital elements.

\section{Survey of $SR(e,i,\omega)$ for selected resonances}
\label{surv}

\begin{figure}
	\centering
	\includegraphics[width=1\linewidth]{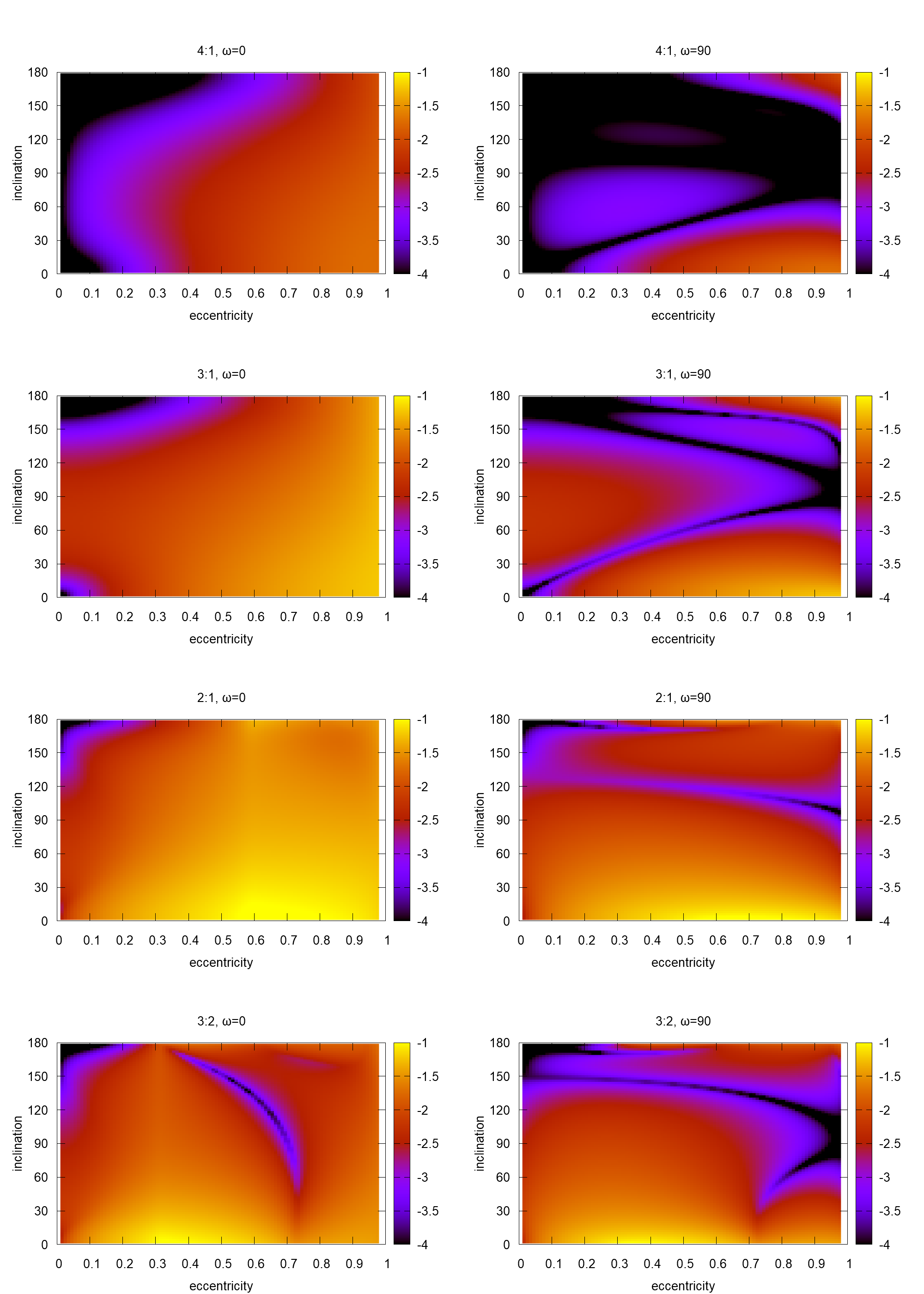}
	\caption{Numerically computed $SR(e,i)$ in log scale corresponding to planet Jupiter for interior resonances 4:1, 3:1, 2:1 and 3:2 for $\omega=0^{\circ}$ at left panels and for  $\omega=90^{\circ}$ at right panels.}
\label{first8}
\end{figure}

\begin{figure}
	\centering
	\includegraphics[width=1\linewidth]{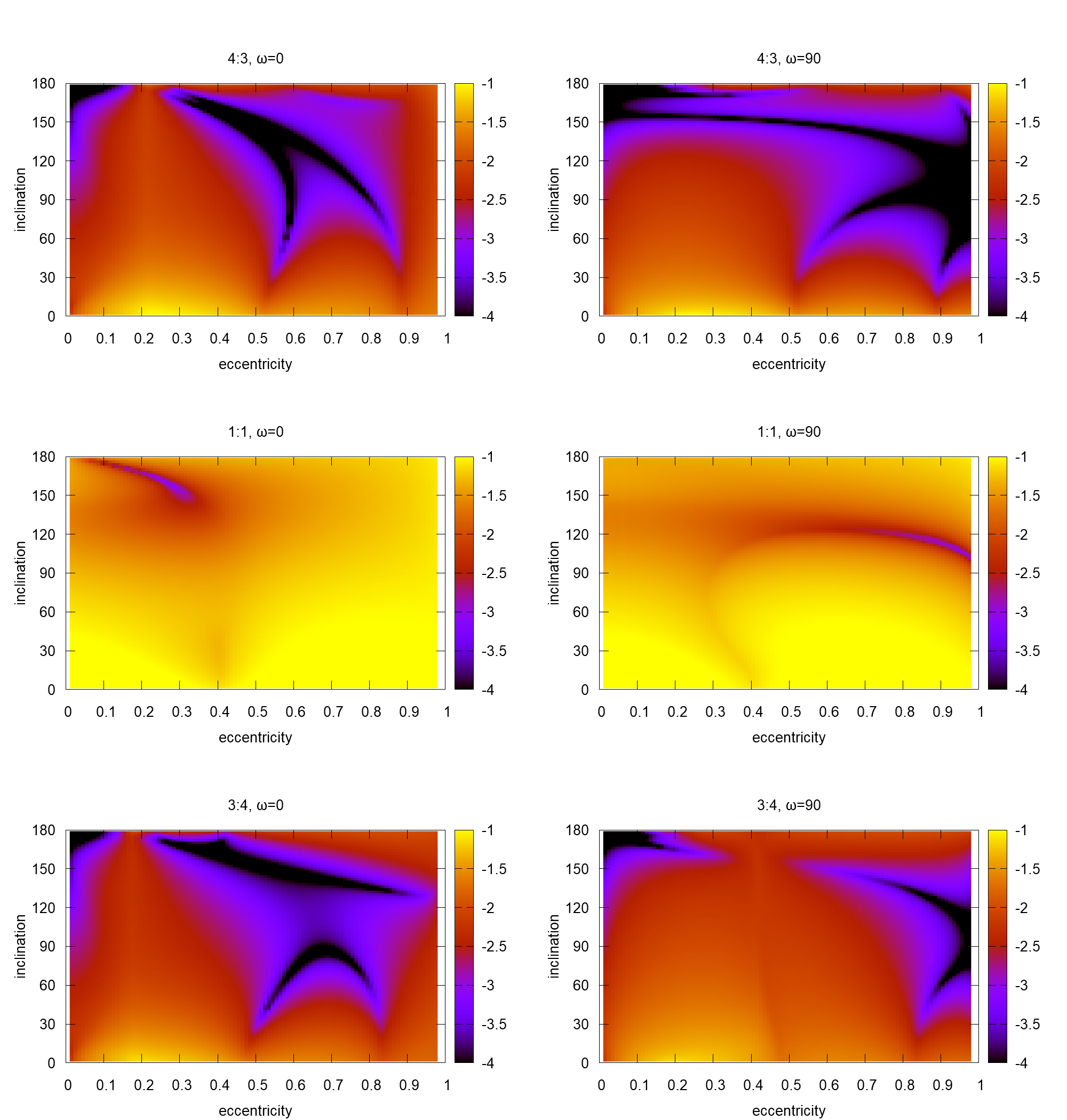}
	\caption{Numerically computed $SR(e,i)$  in log scale corresponding to planet Jupiter for resonances 4:3, 1:1 and 3:4 for $\omega=0^{\circ}$ at left panels and for  $\omega=90^{\circ}$ at right panels.}
\label{secon6}
\end{figure}

\begin{figure}
	\centering
	\includegraphics[width=1\linewidth]{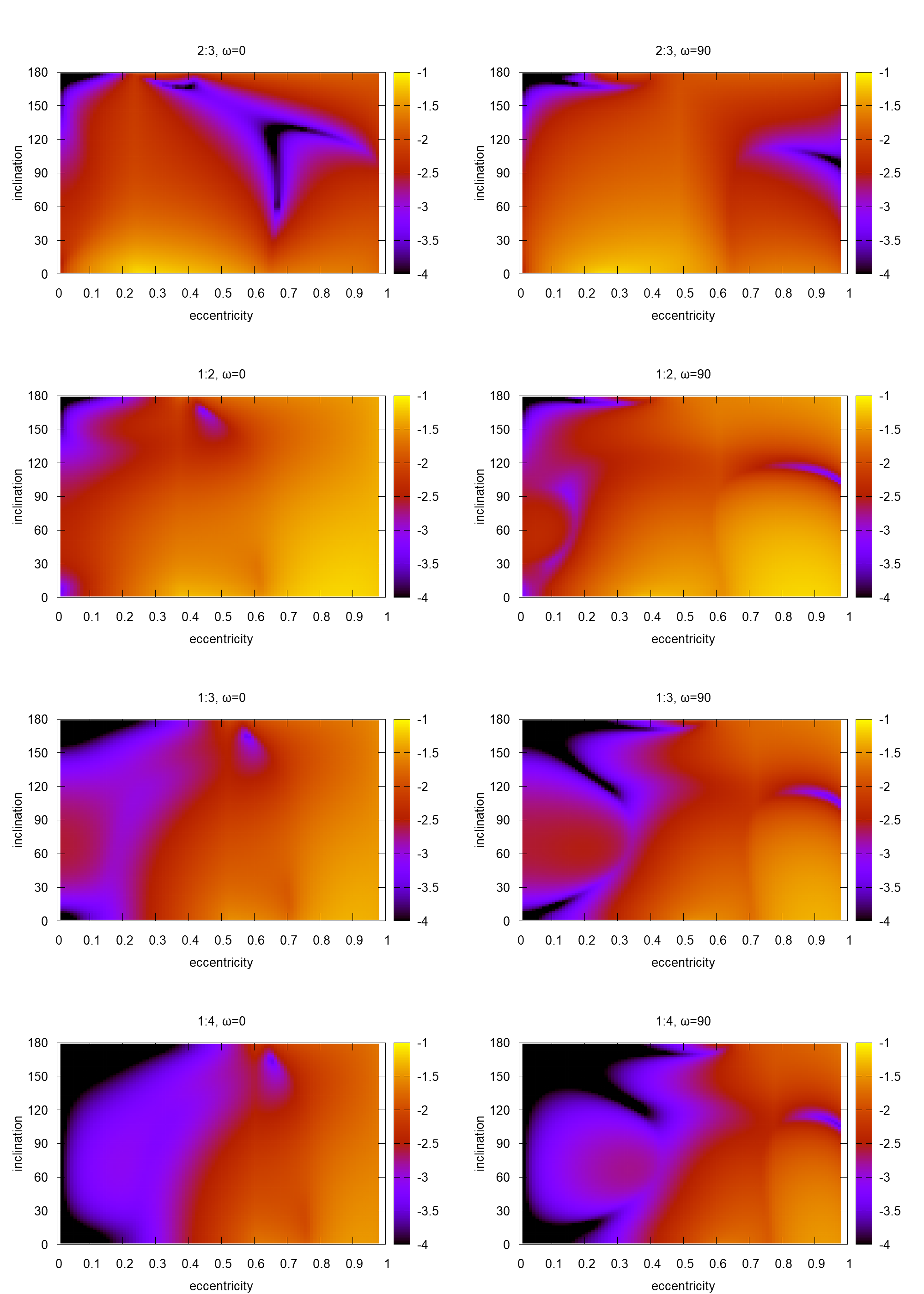}
	\caption{Numerically computed $SR(e,i)$  in log scale corresponding to planet Jupiter for exterior resonances 2:3, 1:2, 1:3 and 1:4 for $\omega=0^{\circ}$ at left panels and for  $\omega=90^{\circ}$ at right panels.}
\label{third8}
\end{figure}

We have computed $SR(e,i,\omega)$ for several  resonances and we can summarize the results analyzing a selection of cases.
Figures \ref{first8} to \ref{third8} show the computed values of the strength $SR(e,i)$ for some internal and external  resonances of order 1, 2 and 3 with Jupiter and also for coorbitals (order 0). They are presented ordered with growing $a$. Jupiter was assumed in circular and with zero inclination orbit. For each resonance we show two plots: one for $\omega=0^{\circ}$ at left and another for $\omega=90^{\circ}$ at right. 
From geometrical arguments, we note that $SR$ is $\pi$-periodic in $\omega$ and symmetric with respect to $\pi/2$.
Bright colors correspond to regions of the plane $(e,i)$ where resonances are strong (which correspond to large widths in the domain of $a$) and dark colors to regions of the plane $(e,i)$ where resonances are weak (narrow widths in the domain of $a$).
The same logarithmic scale was used in all plots to make comparisons more easy.
 These plots are not dynamical maps, they do not show resonance centers or borders, they show variations in $R(\sigma)$ which are related to the widths of the resonances.
Nevertheless, 
we have verified that 
the weak regions in the form of bands that some resonances show in the space $(e,i)$  in general are produced by transitions in the stability of the equilibrium points. In these dark regions the resonance changes its topology and in general a stable equilibrium point turns to be unstable and reciprocally. 
 That can be easily checked calculating $R(\sigma)$ at some points in the figures, as we have done in figure \ref{map1to5} right panel.
Looking at the mosaic of figures we can verify some notable things: 

i) For both
 $\omega=0^{\circ}$ and  $\omega=90^{\circ}$  the dependence of $SR(e,i=0^{\circ})$ is the same because for coplanar orbits the argument of the perihelion is irrelevant. The same occurs for 
  $SR(e,i=180^{\circ})$.

  ii) In each plot it can be verified that 
 $SR(e,i=0^{\circ}) > SR(e,i=180^{\circ})$ but  $dSR(e,i=0^{\circ})/de < dSR(e,i=180^{\circ})/de$, that means, retrograde resonances are weaker and they have a steeper dependence with eccentricity
than direct ones as showed by \cite{2012MNRAS.424...52M} and \cite{2013CeMDA.117..405M}. 

iii) Another expected result is that  $SR(e\sim 0,i)$ is independent of $\omega$ whatever the resonance.

iv) All resonances, with exception of  the exterior resonances of the type 1:$k$, are strongly dependent on $\omega$.
For  $\omega\sim 90^{\circ}$ they almost vanish for inclinations in the interval $60^{\circ} \lesssim i \lesssim 120^{\circ}$ for very high eccentricities constituting a counterintuitive behavior (high eccentricity and drop in strength). Compare for example the resonance 4:1 for $\omega=90^{\circ}$ (figure \ref{first8}) with 1:4 for $\omega=90^{\circ}$ (figure \ref{third8}), or 3:1 with 1:3. This drop of $SR$ with the eccentricity also appear in resonances 1:$k$ but it is minimal and restricted to $i\sim 100^{\circ}$ and $e\rightarrow 1$.

 In particular we can see in figure \ref{first8}  the $SR$ corresponding to the resonance 3:1 that we have analyzed  in previous figures. For the case  $\omega=0^{\circ}$ (left panel) it is evident that $SR$ grows with $e$ but in the case $\omega=90^{\circ}$ (right panel) $SR$ diminishes for growing $e$ for $60^{\circ}\lesssim i \lesssim 120^{\circ}$.
 There is a geometric explanation for this behavior: high-eccentricity near polar orbits with  $\omega\sim 90^{\circ}$ are the ones with greater distances to the planet's orbit and consequently the ones with the weaker resonant perturbations.  That is why for higher eccentricities the resonance strength diminishes, contrary to the classical near coplanar case.
 
 However, we remark that exterior resonances of the type 1:$k$ are the strongest among resonances with neighbor semi-major axes and they have weaker dependences with the orbital inclination especially for high eccentricities (figure \ref{third8}). They are  almost independent of the inclination and the argument of the perihelion being strong even for retrograde orbits. This makes the exterior resonances of the type 1:$k$ more prone to capture eccentric retrograde objects than interior ones as reported for example by \citet{2015MNRAS.446.1998N} and \citet{2016MNRAS.461.3075F}.

Please remember, though, that the orbital elements $(e,i,\omega)$ of the particle will actually evolve on a secular timescale, 
and this could possible bring a resonant particle towards a region where the resonance becomes very weak.
 In this case its semimajor axis will be more easily affected by other perturbations and eventually the resonance could be broken. The dark regions of the $SR(e,i,\omega)$ are the regions where a resonance can be broken more easily, even ignoring close encounters with other planets, and where resonant objects could be found with less probability. On the other hand, the strongest regions in $(e,i)$ of a resonance are in general associated to shorter distances particle-planet and if the critical angle is having large amplitude librations a close encounter can happen disrupting the resonant link. Moreover, it is possible that for that regions in $(e,i)$ other resonances are also strong and wide, generating a resonance overlap which drives the particle to a chaotic evolution.

\subsection{Particular case: Trojans and resonance 1:1 with Jupiter}

\begin{figure}
	\centering
	\includegraphics[width=1.\linewidth]{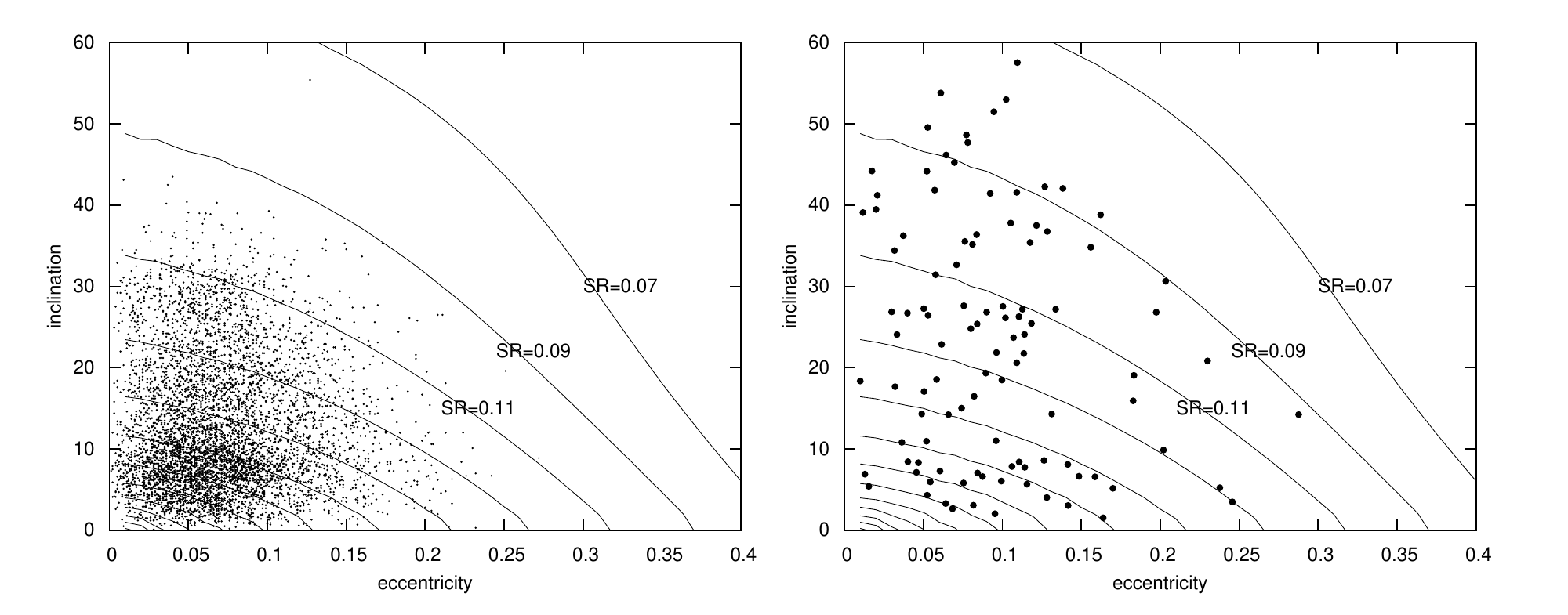}
	\caption{Left panel: mean of left and right panels of figure \ref{secon6} corresponding to resonance 1:1 with Jupiter represented with level curves plus the actual population of 5482 Trojans with multi opposition orbits taken from MPC. These known Trojans are approximately located in a region defined by $SR > 0.09$. Right panel: survivors of a simulation of initial 1:1 resonant fictitious particles after 100 Myr of dynamical evolution. }
\label{trojan}
\end{figure}

In the case of the resonance 1:1, the dependence of $SR$ with $\omega$ in the region $e<0.4$ and $i<60^{\circ}$ is very weak so we computed a mean of $SR$ for $\omega=0^{\circ}$ and $\omega=90^{\circ}$. The result in form of level curves is shown
in figure \ref{trojan} where in the left panel we also plot 
the actual population of objects  classified by MPC as Jupiter's Trojans 
\footnote{www.minorplanetcenter.net/iau/lists/JupiterTrojans.html}. It is interesting to note that they are not located inside a region delimited by $e$ or $i$ but inside a region delimited by an approximately constant value of $SR$. This suggests that outside that region the strength is not enough to sustain a Trojan for a long time interval. 
There is a large list of studies on the stability of Trojans 
but they are mainly referred to the planes $(a,e)$, $(a,i)$,
$(\Delta \sigma,e)$ or $(\Delta \sigma,i)$ and focused in the real population
which is concentrated at low $e$ and $i$ due to cosmogonic reasons. 
In order to test a more uniformly distributed population,
in the right panel we plot the survivors after 100 Myrs of orbital evolution of a population of 1000 fictitious particles originally in coorbital orbits with Jupiter. Their initial conditions were taken at random in the intervals given by $5.15 < a < 5.25$ au, $0 < e < 0.4$,
$0^{\circ}<i<60^{\circ}$ and $(\omega,\Omega,M)$ taken between $0^{\circ}$ and $360^{\circ}$. 
Only the outer planets were considered for this simulation. The 
	survivors are located under the region defined from high-$i$ and small $e$ to low-$i$ and large $e$
 following approximately the level curves of $SR$.
Even considering that there are secular effects that we are not taking into account in our model,
 in this plot it can be recognized a preference of coorbital objects to remain in regions where the resonance is strong.

The object 2015 BZ509 classified as coorbital \citep{2017Natur.543..687W} has $\omega=257^{\circ}$ and for this $\omega$ the function $SR(e,i)$ is approximately similar to the one given in figure \ref{secon6} at central right panel.
This object has orbital elements
 $e=0.38$ and $i=163^{\circ}$, which places it above the relatively weak region approximately delimited by $110^{\circ}<i<140^{\circ}$ and that it is generated by a change in the location and stability of the equilibrium points.
 If its orbital elements remain constant its stability will be assured, but since there is some evolution in $e$, $i$ and specially in $\omega$, it is very likely that at some point it will pass through a weak region where the resonant link can be destroyed. 
 By doing the same reasoning but backwards in time, the object
 could have reached its current position coming from a region where the resonance is unstable
 and then trapped in the present temporarily stable region.

\subsection{Particular case: Hildas and resonance 3:2 with Jupiter}

\begin{figure}
	\centering
	\includegraphics[width=0.9\linewidth]{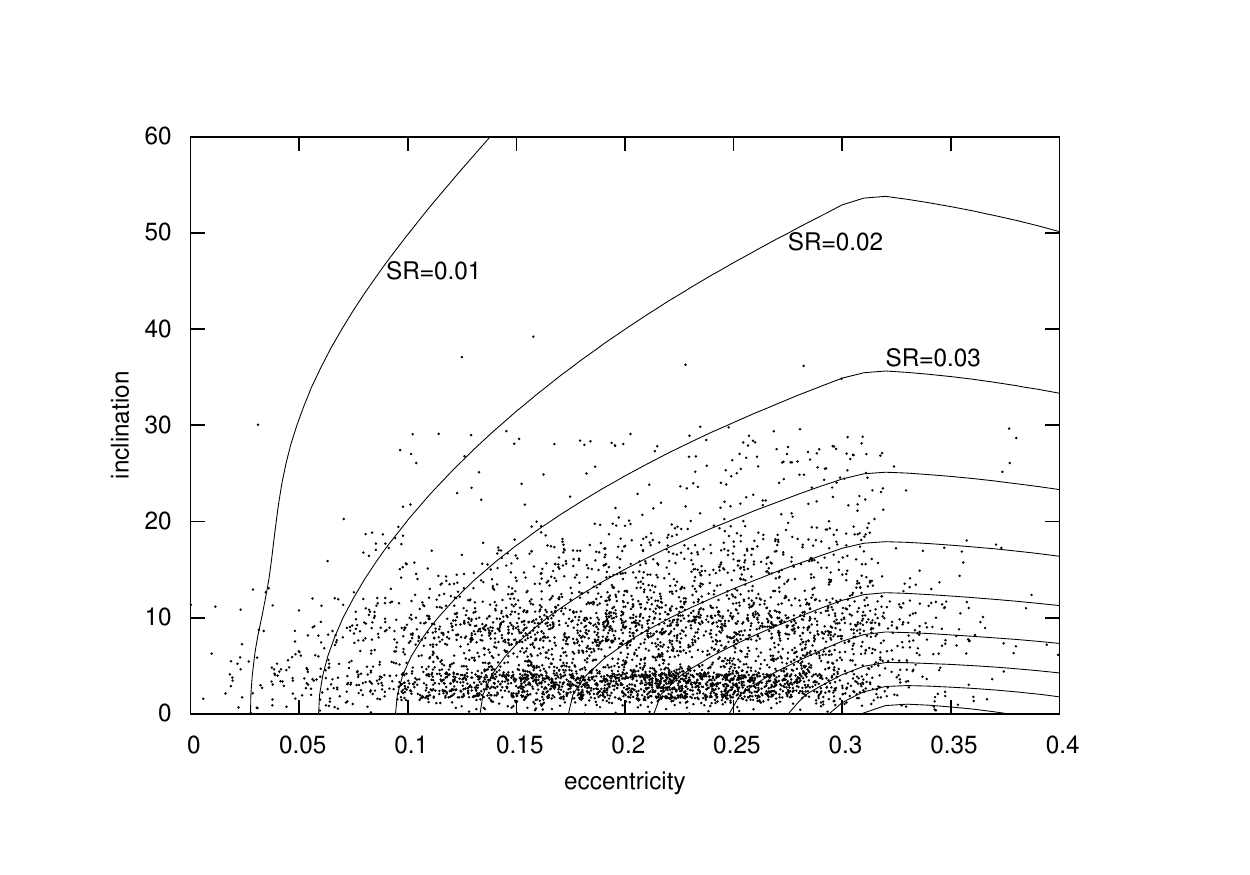}
	\caption{Mean of left and right panels of figure \ref{first8} corresponding to resonance 3:2 with Jupiter represented with level curves plus the actual population of 4487 Hildas  taken from JPL. The Hildas tend to avoid low eccentricity regions where the resonance is weak.  }
	\label{hildas}
\end{figure}

It is interesting to compare Trojans with Hildas because $SR$ has a different behavior in both resonances. While in resonance 1:1 $SR$ grows for lower eccentricities, in the case of the resonance 3:2 $SR$ grows with the eccentricity up to $e \sim 0.32$ and then diminishes. In figure \ref{hildas} 
we show
the mean of $SR$ for $\omega=0^{\circ}$ and $\omega=90^{\circ}$ by means of level curves and
 we also plot 
the actual population of objects with $3.80 < a < 4.05$ au taken from JPL 
\footnote{ssd.jpl.nasa.gov}, which corresponds to the domain of the Hildas. The match is not very good because for cosmogonic reasons there are few high-inclination objects and due to close encounters with Jupiter there are few objects with $e>0.32$. Moreover, the well defined pattern due to the two collisional families identified in this population \citep{2008MNRAS.390..715B} biases the distribution of points in the plot.
Anyway, it is clear that the population is not concentrated at low eccentricities and that the higher the eccentricity the larger the number of Hildas. Also, there are more high-inclination Hildas at large eccentricities than at low eccentricities.

\subsection{Particular case: comet Halley and the resonance 1:6 with Jupiter}

\begin{figure}
	\centering
	\includegraphics[width=0.9\linewidth]{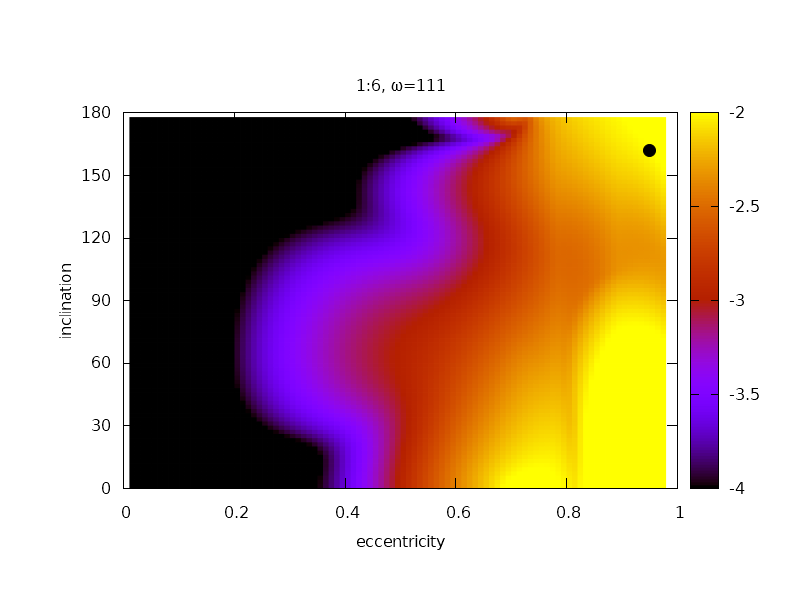}
	\caption{$SR(e,i)$ for resonance 1:6 with Jupiter using Halley's argument of perihelion  $\omega=111^{\circ}$. The black dot indicates the position of an hypothetical  Halley in case it were trapped in this resonance. }
	\label{halley}
\end{figure}

The capture of comets in MMRs with Jupiter is reported in the literature
and there are works that indicate that
 resonances of the type 1:$k$ are preferred \citep{1987A&A...187..899C,1997Icar..125...32C,2005ASSL..328.....F}. For example, more recently
\citet{2016MNRAS.461.3075F} confirmed that in numerical integrations of clones of Halley type comets  there was a remarkable preference for being trapped in resonances 1:$k$ with Jupiter, in particular  in the resonance 1:6 at $a=17.17$ au. We calculated $SR(e,i)$ for this resonance taking $\omega=111^{\circ}$, which is the value corresponding to the comet Halley. Figure \ref{halley} shows the result where we also indicate  the corresponding position 
of an object with the same eccentricity and inclination of
Halley. As we have said, resonances 1:$k$ for high eccentricities are always strong and almost independent of $(i,\omega)$. On the other hand, it is very unlikely that objects with $e<0.5$ could be trapped in this resonance as can be deduced from the computed values of $SR$. But, even if an object is inside a region of $(e,i,\omega)$ where the resonance is strong, its stability strongly depends on the perturbations by the outer planets that we have neglected.

\subsection{Particular case: a polar resonant TNO}

\begin{figure}
	\centering
	\includegraphics[width=1.\linewidth]{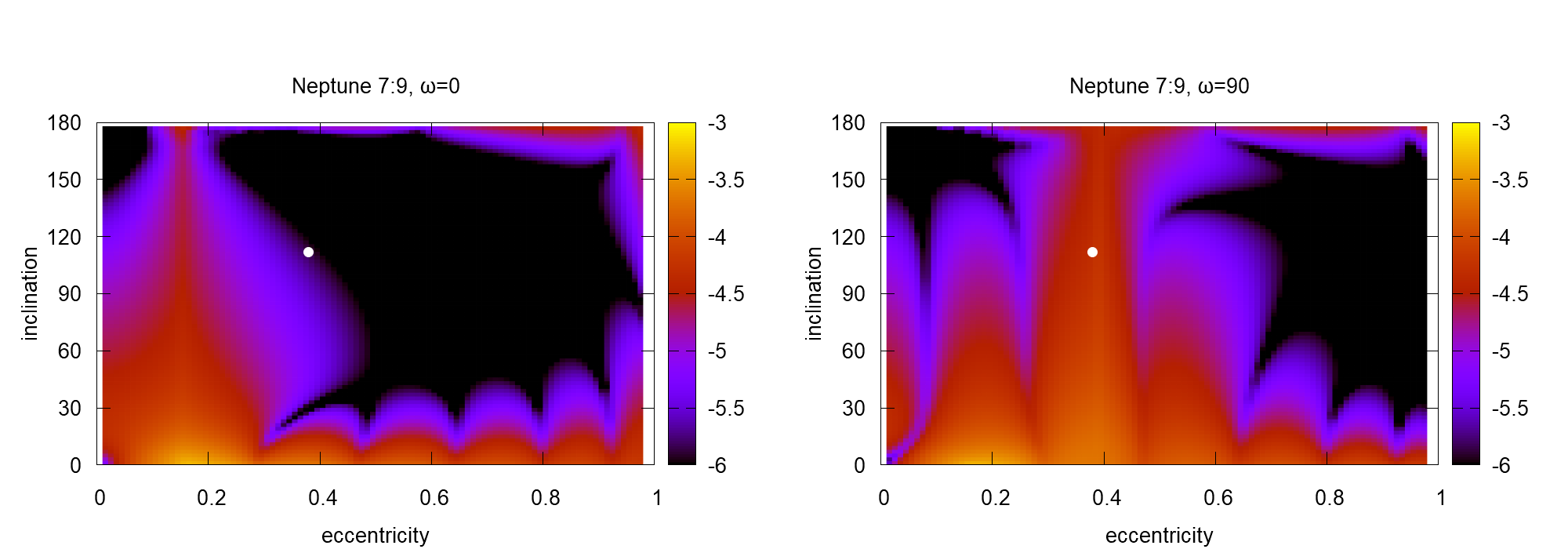}
	\caption{$SR(e,i)$ for resonance 7:9 with Neptune for $\omega=0^{\circ}$ (left) and $\omega=90^{\circ}$ (right). The approximate position of the TNO (471325) is indicated with a white dot.}
\label{polartno}
\end{figure}

\citet{2017MNRAS.472L...1M} presented the very interesting case of the TNO (471325) which is evolving in the exterior 7:9 resonance with Neptune in an almost polar orbit. We calculated two maps of $SR$ for that resonance, one for $\omega=0^{\circ}$ (figure \ref{polartno} left panel)  and another  for $\omega=90^{\circ}$  (figure \ref{polartno} right panel). 
A white dot indicates the present values of $e,i$ for this object which do not vary much in the resonant timescale. 
According to \citet{2017MNRAS.472L...1M} the object has a circulating $\omega$ and it 
is interesting to note that when $\omega=0^{\circ}$ or $180^{\circ}$ the libration amplitude  of $a$ is small, and 
 when
 $\omega=90^{\circ}$ or $270^{\circ}$, the libration amplitude  of $a$ is large. This is in agreement with figure \ref{polartno} that shows that the object is inside a weak region of the resonance when $\omega=0^{\circ}$ and inside a strong region  when $\omega=90^{\circ}$.

\section{Resonant structure in the $(a,i)$ plane}
\label{strucai}

\begin{figure}
	\centering
	\includegraphics[width=1.\linewidth]{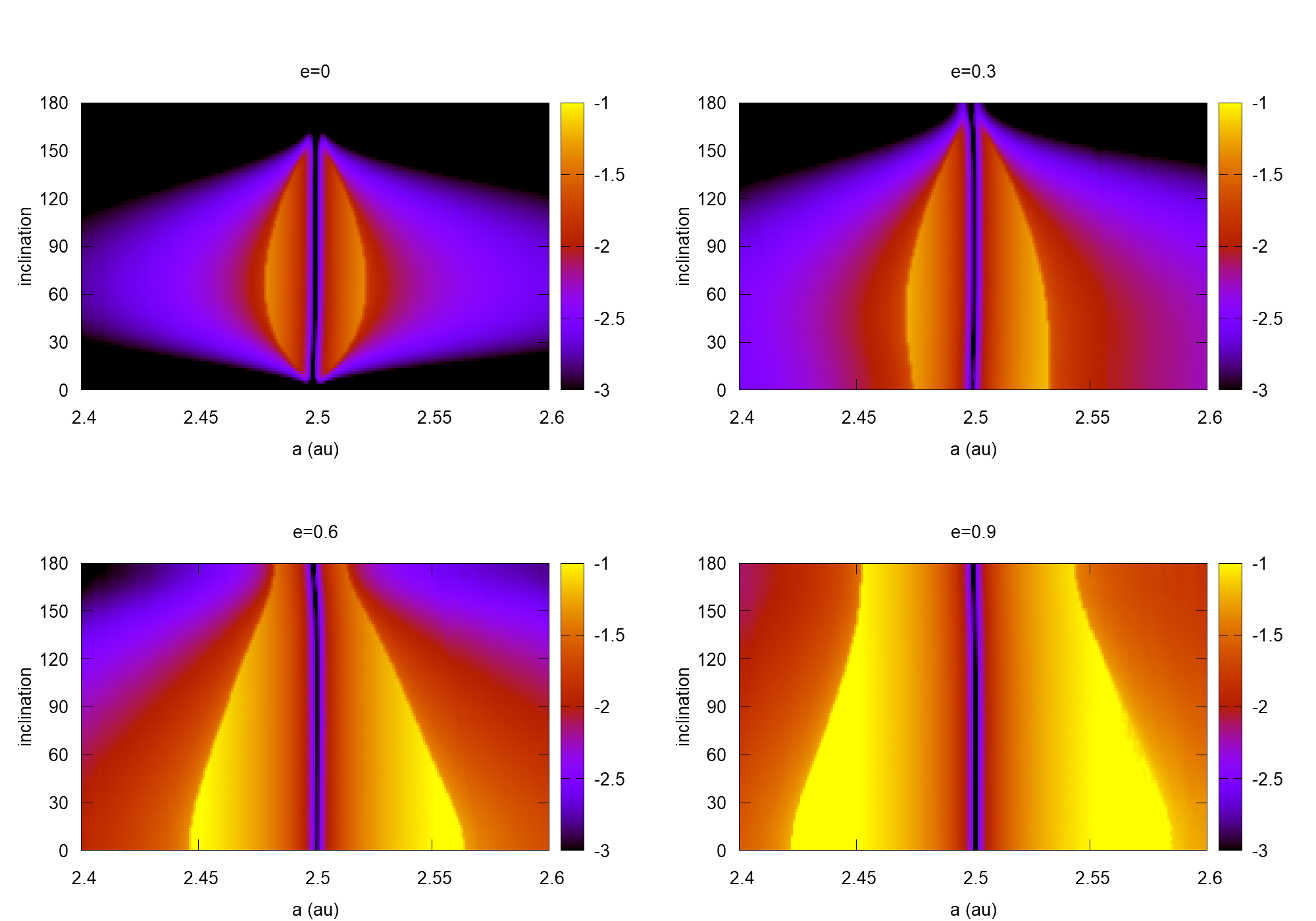}
	\caption{Dynamical maps of the interval of variations $\Delta a$ in logarithmic scale showing the structure of resonance 3:1 in the space $(a,i)$ for 4 different eccentricities assuming $\omega=0^{\circ}$ and only Jupiter as perturbing planet. The center of the resonance is the dark vertical line defined in all frames at $a=2.50$ au, although it is possible to see a slight departure form that nominal value for $i>150^{\circ}$. The limits of the resonance, where the separatrices are located,
	are the sharp curves limiting the yellow zone from orange zone.}
\label{3to1}
\end{figure}

We have showed that 
the numerically computed function $SR(e,i,\omega)$ provides a good description of the resonance strength and 
by extension an
indication of the resonance width.
Nevertheless, $SR$ by itself cannot provide details of the structure in the space $(a,e,i)$ where librations take place and where equilibrium points and separatrices are located. For this kind of details is necessary to analyze
plots of the level curves of $\mathcal{K}$ in the plane $(a,\sigma)$, to
apply analytical methods when possible 
\citep{2018MNRAS.474..157N}
or to explore the resonance by mean of dynamical maps as we have explained in section \ref{map}.

We are used to the shape of the resonances in the plane $(a,e)$ for near zero inclination orbits (see for example figure 2 in \cite{2002aste.conf..379N}) but 
very few studies have been done in the space $(a,i)$ \citep{2010P&SS...58.1906C,2015MNRAS.451.2399C,2018MNRAS.474..157N}. 
In figure \ref{3to1} we show four dynamical maps of $\Delta a$ in the plane $(a,i)$ obtained for the resonance 3:1 with Jupiter assuming $\omega=0^{\circ}$, each one corresponding to a given eccentricity for the test particles. Each plot corresponds to a vertical line in figure \ref{first8} second row left panel for $e=0$, $e=0.3$, $e=0.6$ and $e=0.9$ respectively.

In the first map, for $e=0$, the resonance width is maximum for  $i\sim 60^{\circ}$ in agreement with the $SR$ showed in \ref{first8}, and vanishes for $i\sim 0^{\circ}$ and  $i\sim 180^{\circ}$. In the second map for $e=0.3$ the resonance width is almost constant from $i\lesssim 60^{\circ}$ and then diminishes in agreement with the $SR(e)$ that one can infer from figure  \ref{first8}. This map for $e=0.3$ can be compared with figure 2 in \citet{2018MNRAS.474..157N} where there is a perfect match for $i\sim 0^{\circ}$, but in their figure even adding both contributions, retrograde and direct, the total width that our dynamical map shows for $i>0^{\circ}$ is not attained. As we have explained earlier, the analytical widths obtained by \citet{2018MNRAS.474..157N} are the ones associated to specific resonant terms while in the full numerical integrations of the dynamical maps and in our computed $SR$ all terms contribute to the final result. This confirms the necessity to consider the whole disturbing function whenever we depart from the central value of the expansion used.
In the last two maps the width diminishes for growing inclinations in agreement to what can be deduced from the $SR$ in figure  \ref{first8}. 
 The maps computed for other values of  $\omega$ can change dramatically as can be deduced from figure   \ref{first8}  right panel.

The dynamical maps we have showed were obtained considering   only the planet Jupiter as a perturber and in circular  and zero inclination orbit. 
When considering the actual planetary system the general properties of the maps for the interior resonances with Jupiter persist but new structures can show up due to secular effects or MMRs with other planets and their overlaps \citep{2002mcma.book.....M}. In particular, the resonances located between the giant planets are strongly affected. The situation is worst
for near coplanar  and high eccentricity orbits  because close encounters particle-planet are more likely and resonances can only survive in some stable regions.

\section{Conclusions}
\label{conc}

In the dynamics of small bodies perturbed by one planet on a circular orbit, the geometry of any MMR generally depends strongly on both $e$ and $i$, as well as on 
$\omega$ since the resonance cannot be represented by a single critical angle. The conventional picture of a main resonant angle comes from the
formalism of series expansions, and it does not hold in the overall space of orbital elements. However, the structure of MMRs as a function of $(e,i,\omega)$
can still be explored in a unified way by using the whole perturbing function computed numerically.
In particular, the numerical function $SR(e,i,\omega)$ is a measure of the strength (that is the depth) of the resonances, and, by extension
it is also a good indicator for their  width in the whole domain $(e,i,\omega)$.
Indeed, for high inclination and/or high eccentricity orbits the total width of a resonance is not the dynamical result of an isolated resonant term in the disturbing function but the result of the full disturbing function that must be computed numerically. In this context the proposed numerically computed strength $SR$  proves to be very useful and, in the examples we have analyzed, $\sqrt{SR}$ is approximately proportional to the resonance width.

Contrary to what series expansions show in restricted regions of the orbital elements space, the function $SR$ proves that the strength and width of MMRs are not monotonic functions of $e$ and $i$. The strength and width even drop abruptly 
in some regions of the space $(e,i,\omega)$, meaning that the resonance itself nearly vanish.
Exterior resonances of the type 1:$k$ are particularly strong for high eccentricities independently of $(i,\omega)$. But all other resonances have the characteristic
 that for
 $60^{\circ}\lesssim i \lesssim 120^{\circ}$
and $\omega$ near $90^{\circ}$ or  $270^{\circ}$ the strengths and widths drop greatly for growing eccentricities at high eccentricity regime.

Regions where $SR$ is very low represent regions where the resonance can be broken more easily by  perturbations from other planets of the Solar System. 
However, a high value of $SR$ does not absolutely guarantee that a regular resonant motion is possible. Indeed, the width of neighbor resonances could be large as well, creating a chaotic region due to resonance overlap.
On the other hand, a close encounter with a planet could abruptly break the resonant motion whatever the value of $SR$. 

The structure of resonances in the space $(a,i)$ has analogue elements with the structure in the space $(a,e)$. In particular, each resonance has some range of inclinations for which  
zero-eccentricity resonant orbits are possible, contrary to the classic paradigm coming from near-zero inclination theories.

\textbf{Acknowledgments.}
I am indebted to two reviewers who contributed greatly to improve the original version of the manuscript.
I acknowledge support from the Comisi\'on Sectorial de Investigaci\'on Cient\'ifica
(CSIC) of the University of the Republic through the project CSIC Grupo I+D
831725 - Planetary Sciences and PEDECIBA.

\appendix
\section{Hamiltonian and disturbing function}
The Hamiltonian describing the resonant dynamics is \citep{1998CeMDA..71...35M,2002aste.conf..379N}:
\begin{equation}\label{ham}
\mathcal{K}(a,\sigma) = -\frac{\mu}{2a} -n_p\frac{k_p}{k}\sqrt{\mu a}  - R(a,\sigma)
\end{equation}
where $\mu = G m_{\odot}$ is the gravitational parameter of the Sun.
The study of the level curves of $\mathcal{K}(a,\sigma)$ gives us 
a description of the resonant motion with the location of the equilibrium points, the form of the islands of oscillations around the stable equilibrium points and the location of the separatrices passing by the unstable equilibrium points limiting the largest oscillations.
Using a simplified resonant disturbing function of the form $R = A\cos(\sigma)$, introducing this expression in Eq. (\ref{ham}) and using a Taylor development around the resonance centre $a_0$ we get a pendulum Hamiltonian, called \textit{first fundamental model of resonance} by \citet{1983CeMec..30..197H}. But it is also possible to calculate 
$R(a,\sigma)$ numerically avoiding the limitations of the simplified models and then, to obtain the exact value of $\mathcal{K}(a,\sigma)$.

To obtain numerically 
the resonant disturbing function $R(a,\sigma)$ corresponding to the resonance 
$k_p:k$
in the interval $0^{\circ}\leq \sigma \leq 360^{\circ}$, following \citet{2006Icar..184...29G} we have to calculate:

\begin{equation}\label{rmean}
R(a,\sigma)=\frac{1}{2\pi k}\int_{0}^{2\pi k}\mathbb{R}(\lambda_p,\lambda(\lambda_p,\sigma))d\lambda_p
\end{equation}
being $\mathbb{R}$ the planetary disturbing function 
\begin{equation}\label{defr}
\mathbb{R}=G m_p\Bigl(\frac{1}{\mid \mathbf{r}_p-\mathbf{r}\mid} - \frac{\mathbf{r}\cdot\mathbf{r}_p}{r^3_p}
\Bigr)
\end{equation}
for a given set of fixed values of $(e,i,\omega,\sigma)$ where we have expressed
$\lambda=\lambda(\lambda_p,\sigma)$ from (\ref{sigma}) with $\sigma$ as a fixed parameter and where
$\mathbb{R}(\lambda_p,\lambda)$ is evaluated numerically from (\ref{defr}) where the heliocentric position vectors $\mathbf{r}_p$ and $\mathbf{r}$
were expressed as functions of the orbital elements and mean longitudes $\lambda_p$ and $\lambda$. 
The integral is computed in the interval $0\leq \lambda_p \leq 2\pi k$ which is the variation period of domain of $R$.
We repeat for a series of values of $\sigma$
between $(0^{\circ},360^{\circ})$ obtaining a numerical representation of the resonant disturbing function
$R(a,\sigma)$.
For the calculation of the integral we assume $a=a_0$ which is the nominal value of the semi-major axis of the resonant particle. 
In the following, we will impose $a=a_0$ and write $R(\sigma)$
as a shortcut for $R(a_0,\sigma)$.
All along this paper $R(\sigma)$ corresponds to a section of $\mathcal{K}(a,\sigma)$ 
 for $a=a_0$ and there is a correspondence between level curves of constant $\mathcal{K}$ 
 and  $R(\sigma)$. Illustrative examples for such curves can be found in figure 1 by \citet{2006CosRe..44..440S}.

\bibliographystyle{elsarticle-harv}
\bibliography{bibliospace}

\end{document}